\documentclass[12pt,letterpaper]{article}
\usepackage[a4paper, total={6.6in, 10in}]{geometry}

\usepackage{helvet}
\usepackage{authblk}
\usepackage{hyperref}
\usepackage{amsmath}
\usepackage{float}
\usepackage{graphicx}
\usepackage[dvipsnames]{xcolor}
\usepackage[utf8]{inputenc}
\usepackage[utf8]{inputenc}

\makeatletter
\renewcommand{\maketitle}{\bgroup\setlength{\parindent}{0pt}
\begin{flushleft}
  \textbf{\@title}
  
  \@author
\end{flushleft}\egroup}
\makeatother

\title{Optimizing Prompt Strategies for SAM: Advancing lesion Segmentation Across Diverse Medical Imaging Modalities}

\author[1,**]{Yuli Wang}
\author[2,**]{Victoria Shi}
\author[2,3]{Wen-Chi Hsu}
\author[2]{Yuwei Dai}
\author[2]{Sophie Yao}
\author[5]{Zhusi Zhong}
\author[4]{Zishu Zhang}
\author[4]{Jing Wu}
\author[5]{Aaron Maxwell}
\author[5]{Scott Collins}
\author[5]{Zhicheng Jiao}
\author[2,*]{Harrison X. Bai}

\affil[1]{Department of Biomedical Engineering, Johns Hopkins University, Baltimore, MD, USA}
\affil[2]{Department of Radiology and Radiological Science, Johns Hopkins University, Baltimore USA}
\affil[3]{Department of Medical Imaging and Intervention, Chang Gung Memorial Hospital at Linkou, Taoyuan, Taiwan}
\affil[4]{Department of Radiology, Second Xiangya Hospital, Central South University, Changsha, Hunan, China}
\affil[5]{Department of Diagnostic Imaging, Brown University Health, Providence, RI, USA}
\affil[*]{Department of Radiology and Radiological Science, Johns Hopkins University, Baltimore USA}
\affil[**]{Authors contribute equally to this paper}

\usepackage[super,comma,sort&compress]{natbib}

\begin{document}

\maketitle


\section*{Summary}
Optimizing prompt strategies for the Segment Anything Model are explored, focusing on prompt location, number, and reinforcement learning-based agent for prompt placement across four lesion datasets.

\section*{Key Points}
\begin{itemize}
    \item This study is among the first to identify optimal thresholds for SAM prompt strategies, tailored to specific pathologies and imaging modalities, providing actionable insights into SAM's adaptability across diverse imaging contexts.
    \item A reinforcement learning-based agent is introduced to reduce decision-making time in SAM-assisted segmentation, advancing toward fully automatic, lesion-specific image segmentation and enhancing its clinical applicability for precision diagnostics and workload reduction.
\end{itemize}

\section*{Abstract}

\textbf{Purpose:} To evaluate various Segmental Anything Model (SAM) prompt strategies across four lesions datasets and to subsequently develop a reinforcement learning (RL) agent to optimize SAM prompt placement.
\\
\textbf{Materials and Methods:} This retrospective study included patients with four independent ovarian, lung, renal, and breast tumor datasets. Manual segmentation and SAM-assisted segmentation were performed for all lesions. A RL model was developed to predict and select SAM points to maximize segmentation performance. Statistical analysis of segmentation was conducted using pairwise t-tests.\\
\textbf{Results:} Results show that increasing the number of prompt points significantly improves segmentation accuracy, with Dice coefficients rising from 0.272 for a single point to 0.806 for five or more points in ovarian tumors. The prompt location also influenced performance, with surface and union-based prompts outperforming center-based prompts, achieving mean Dice coefficients of 0.604 and 0.724 for ovarian and breast tumors, respectively. The RL agent achieved a peak Dice coefficient of 0.595 for ovarian tumors, outperforming random and alternative RL strategies. Additionally, it significantly reduced segmentation time, achieving a nearly 10-fold improvement compared to manual methods using SAM. \\
\textbf{Conclusion:} While increased SAM prompts and non-centered prompts generally improved segmentation accuracy, each pathology and modality has specific optimal thresholds and placement strategies. Our RL agent achieved superior performance compared to other agents while achieving a significant reduction in segmentation time.

\section*{Keywords}

Segment Anything Model, Reinforcement Learning, Lesion Segmentation, Decision Making

\section*{Introduction}

Medical image segmentation plays a vital role in patient care by enabling accurate diagnosis, effective treatment planning, and ongoing monitoring of disease progression. Manual segmentation of lesions, anatomical structures, or regions of interest (ROIs) is often a tedious, labor-intensive, and time-consuming process. To alleviate these burdens, significant research efforts have focused on developing deep-learning models for automatic or semi-automatic segmentation. Currently fully automated tools, such as TotalSegmentator \cite{doi:10.1148/ryai.230024}, excel at segmenting standard anatomical structures. However, these models are primarily optimized for normal anatomy and lack the capability to accurately identify pathological regions, such as lesions. Therefore, advancing lesion segmentation in radiological images is essential, as it requires specialized approaches to address the irregular and variable nature of pathological findings. Recent foundational models like the Segment Anything Model (SAM) and its adaptations have shown substantial promise in radiology, demonstrating generalizable performance across diverse imaging modalities and object types while enhancing workflow efficiency \cite{kirillov2023segment,MAZUROWSKI2023102918}.

Previous studies exploring SAM applications in radiology have primarily focused on assessing and optimizing its performance across various datasets, encompassing different ROIs and imaging modalities \cite{MAZUROWSKI2023102918,he2023adaptingsegmentmodelsam,cheng2023sammedicalimagescomprehensive, huang2024segment,yan2022sam,de2023testing,wang2024car,wang2023deep,wang2021development}. While SAM’s unmodified performance varies significantly depending on dataset characteristics such as ROI size, shape, contrast, and imaging modality, it demonstrates improved generalizability with additional fine-tuning \cite{zhang2024sam,ma2024sam,huang2024segment}. Some studies have also explored the effects of different prompt strategies, showing that box prompts and an increased number of prompt points yield better outcomes \cite{cheng2023sam,mazurowski2023segment,huang2024segment,stein2023past,ranem2023exploring}. However, there is still limited research on optimizing SAM prompt strategies tailored to specific ROI characteristics.

Integrating SAM-assisted segmentation into clinical workflow effectively requires careful consideration. As an interactive model, SAM involves sequential decision-making to determine the placement of additional prompts, which can diminish its potential efficiency gains. Approaches like reinforcement learning (RL) could address this challenge by predicting optimal prompt placement and defining stop criteria, potentially reducing decision-making time \cite{huang2024optimizing,hu2023reinforcement}. RL agents mimic the human trial-and-error process to optimize outcomes. Recent work by Wang et al. 2024 \cite{huang2024alignsam} introduced AlignSAM, a framework leveraging RL-predicted prompt positions to enhance segmentation performance, demonstrating improved results. However, further investigation is needed to explore integrating RL agents into SAM segmentation processes, particularly regarding their role in improving efficiency in clinical settings, particularly regarding decision-making speed.

In this study, we examine the number and location of positive prompt points across four diverse datasets, each representing different imaging modalities, lesion types, or organs with distinct ROI characteristics. Based on these insights, we propose a proof-of-concept RL agent aimed at reducing decision-making time in SAM-assisted segmentation. This approach marks a step toward fully automatic, lesion-specific image segmentation using SAM, paving the way for more efficient patient care.

\section*{Methods and Materials}

\subsection*{Patients and Data Collection}
To ensure a diverse representation of lesions across different sizes and imaging modalities, we utilized four independent institutional or publicly available datasets: 1) ovarian tumor, 2) renal tumor, 3) lung tumor, and 4) breast tumor. The lung and breast tumor datasets were sourced from publicly available datasets from The Cancer Imaging Archive (TCIA) (Table \ref{tab:tumor_dataset_summary}).

Inclusion criteria for all datasets were as follows: (1) histopathologically confirmed lesions; (2) pretreatment imaging scans; (3) visible lesions on the respective imaging modality; (4) availability of both T1-weighted contrast-enhanced imaging (T1CE) and T2-weighted imaging (T2WI) for MRI datasets, thin slice thickness (0.75mm) for CT images and 2D slices for ultrasound images. Patients were excluded if their images were of poor quality, failed pre-processing, or could not be segmented using SAM on 3D Slicer. Lesion size and imaging characteristics were recorded for all included cases.

\subsection*{Segmentation} 
\begin{figure}[H]
  \centering
  \includegraphics[width=0.85\textwidth]{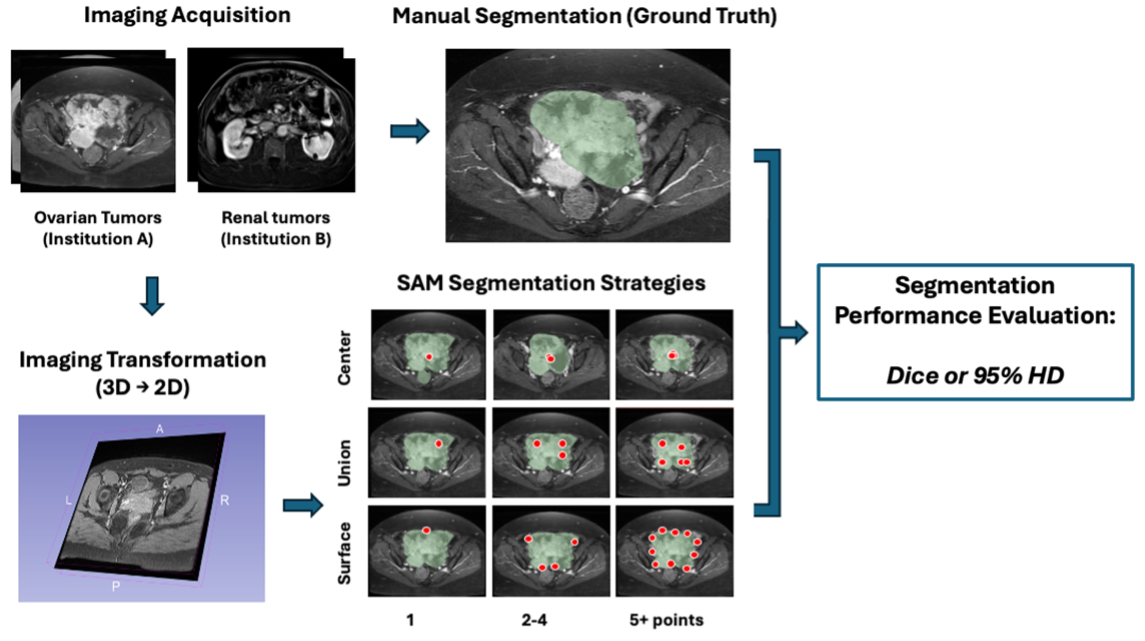}
  \caption{Workflow illustrating SAM-assisted segmentation across all four datasets, showcasing variations in prompt numbers and locations, using the ovarian and renal dataset as an example.}
  \label{fig:SAM_workflow}
\end{figure}

All four primary datasets underwent two segmentation processes: manual ground truth segmentation and semi-automatic SAM-assisted segmentation (Figure \ref{fig:SAM_workflow}). For manual segmentation, two radiologists with more than 5 years of experience delineated the regions of interest using 3D Slicer, which served as the ground truth mask. The region of interest was defined as the largest continuous tumor for all datasets. For SAM-assisted segmentation, three radiologists and three medical trainees used SAM integrated with 3D Slicer, employing a combination of point number and location strategies. The number of prompt points was evaluated in three groups: a single point, 2-4 pints, and 5 or more points. To assess point location, three sub-regions were defined: central (within 5 pixels of the lesion center), surface (within 5 pixels of the lesion edge), and union (the remaining lesion area) (Figure \ref{fig:SAM_workflow}).


\subsection*{RL Agent Training and Architecture}
Our goal is to select a small number of regions (cropped from the original images) as prompts to maximize SAM’s segmentation performance. We frame this task as a sequential decision-making process and train an RL agent to interact in a human-like, interactive manner (illustrated in Figure \ref{fig:rl_workflow}). The agent is designed to engage with the visual features or the kinematic prompt sequence, which constitutes the environment. At each time step, the agent observes the current state of the environment and selects an action (i.e., the next prompt location), which subsequently alters the environment. The agent receives a numerical reward based on its action and updates its policy to maximize cumulative future rewards. In line with the standard reinforcement learning paradigm, we formalize this task as a Markov Decision Process (MDP), drawing inspiration from prior works such Fang et al. \cite{fang2017learning} and Padmakumar et al. \cite{padmakumar2018learning}. Detailed specifications of the action space, state representation, and reward structure are provided in the Supplementary Material, along with a comprehensive description of the RL framework and code.

The prompts selected by the RL agent are used as inputs for SAM, simulating a radiologist's interactive approach. We compare segmentation performance and time using prompts chosen by the RL agent, radiologists, and random selection.


\subsection*{Statistical Analysis}
All statistical analyses were performed using Python with SciPy and Scikit-learn libraries~\cite{scikit-learn}. The ROI sizes were compared using the Mann-Whitney U test. Pairwise t-tests were used to evaluate differences in Dice coefficients and 95th percentile Hausdorff Distance (HD95) values across varying point numbers and sub-region strategies for each dataset. The same metrics (Dice coefficients and HD95) was also analyzed to assess the performance of the RL agent.

\section*{Results}

\subsection*{Patient Characteristics}
Detailed patient characteristics are presented in Table \ref{tab:tumor_dataset_summary}. The mean ROI size of the ovarian tumor, renal tumor, lung tumor, and breast tumor datasets was 8.9, 3.4, 3.0, 17.6, and 3.7 centimeters, respectively.


\subsection*{Segmentation Strategy Performance}
\begin{figure}[H]
  \centering
  \includegraphics[width=0.7\textwidth]{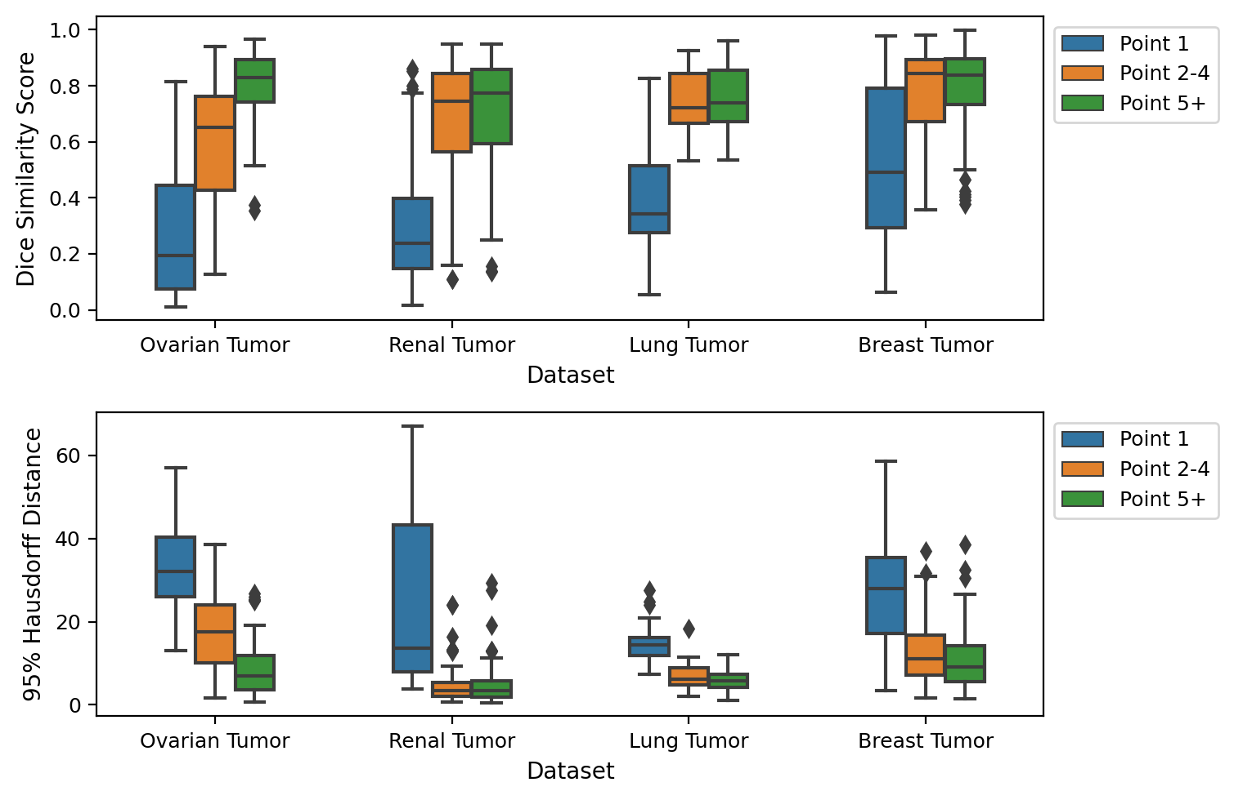}
  \caption{Box plot on Dice Coefficients and 95th Percentile Hausdorff Distance (HD95) across all datasets for various point number strategies.}
  \label{fig:point_number_dice_hd}
\end{figure}

\begin{figure}[H]
  \centering
  \includegraphics[width=0.7\textwidth]{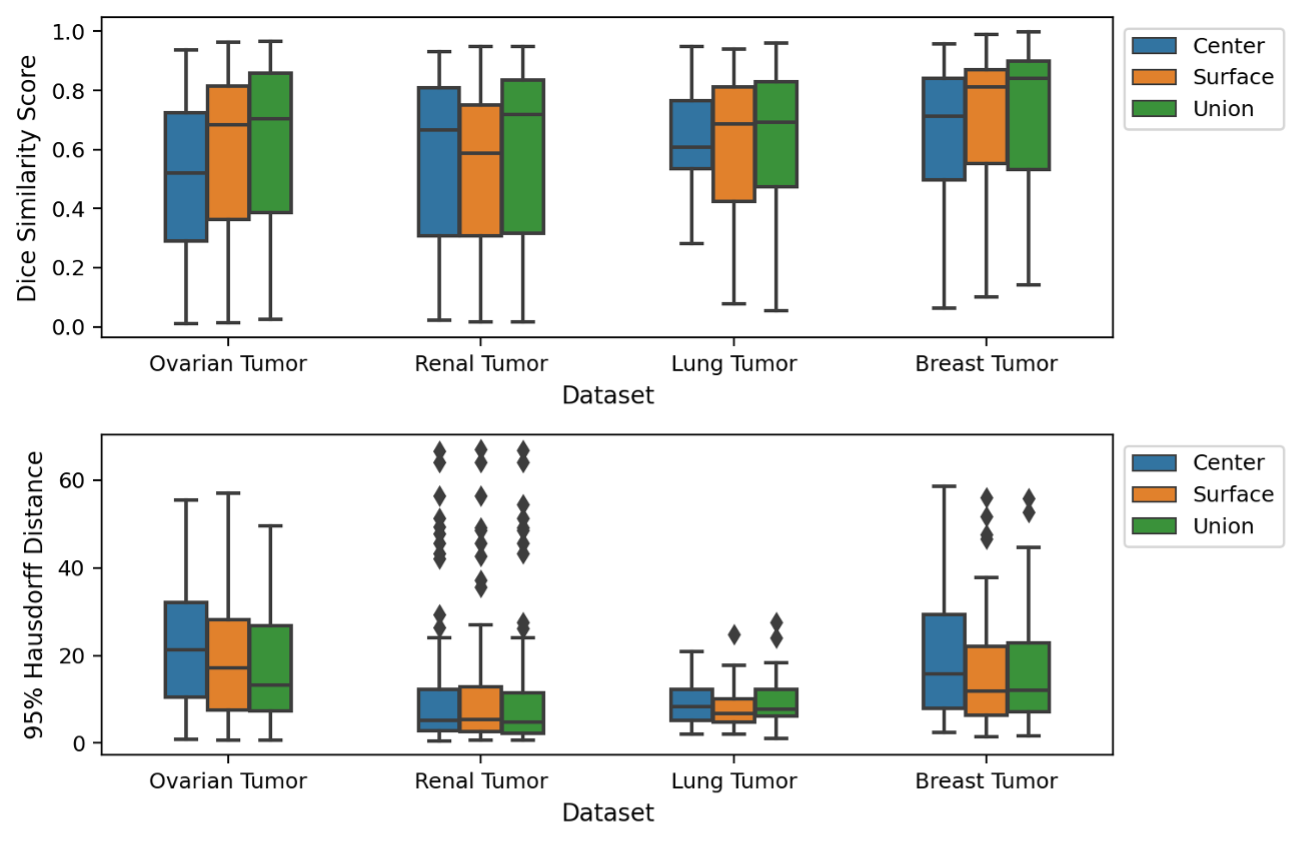}
  \caption{Box plot on Dice Coefficients and 95th Percentile Hausdorff Distance (HD95) across all datasets for various point location strategies.}
  \label{fig:point_location_dice_hd}
\end{figure}

Using the manual segmentation as the ground truth, SAM’s performance across all datasets based on point number strategies is presented in Table \ref{tab:performance_metrics} and Figure \ref{fig:point_number_dice_hd}, while the performance based on point location strategies is detailed in Table \ref{tab:point_location_metrics} and Figure \ref{fig:point_location_dice_hd}. 

\subsection*{Point Strategies}
For the ovarian tumor datasets, the Dice coefficient increased, and HD95 values decreased as the number of prompt points increased, with significant differences found between any two pairs (p = 0.012). 

In the renal tumor, lung tumor, and breast tumor datasets, the Dice coefficient improved initially but plateaued as the number of prompt points increased, showing no significant difference between the 2-4 points group and the 5 or more points groups (p = 0.165). Similarly, HD95 values decreased initially but reached a plateau, with no significant difference between the 2-4 points and 5 or more points groups (p = 0.071). 



\subsection*{Location Strategies}
For the ovarian and breast tumor datasets, the Dice coefficient was significantly higher for the surface and union prompt groups compared to the center group (p $<$ 0.05). Although the union groups showed higher mean Dice coefficients than the surface groups (0.604 vs. 0.580 for ovarian tumors, 0.724 vs. 0.708 for breast tumors), these differences were not statistically significant (p = 0.496 and p= 0.872, respectively). HD95 values were significantly lower for the surface and union groups compared to the center group in both the ovarian tumor and breast tumor datasets (p $<$ 0.05). 

In contrast, for the lung tumor and renal tumor datasets, there were no significant differences in Dice coefficient and HD95 values among the point location groups.



\subsection*{RL Agent Performance}
\begin{figure}[H]
  \centering
  \includegraphics[width=0.9\textwidth]{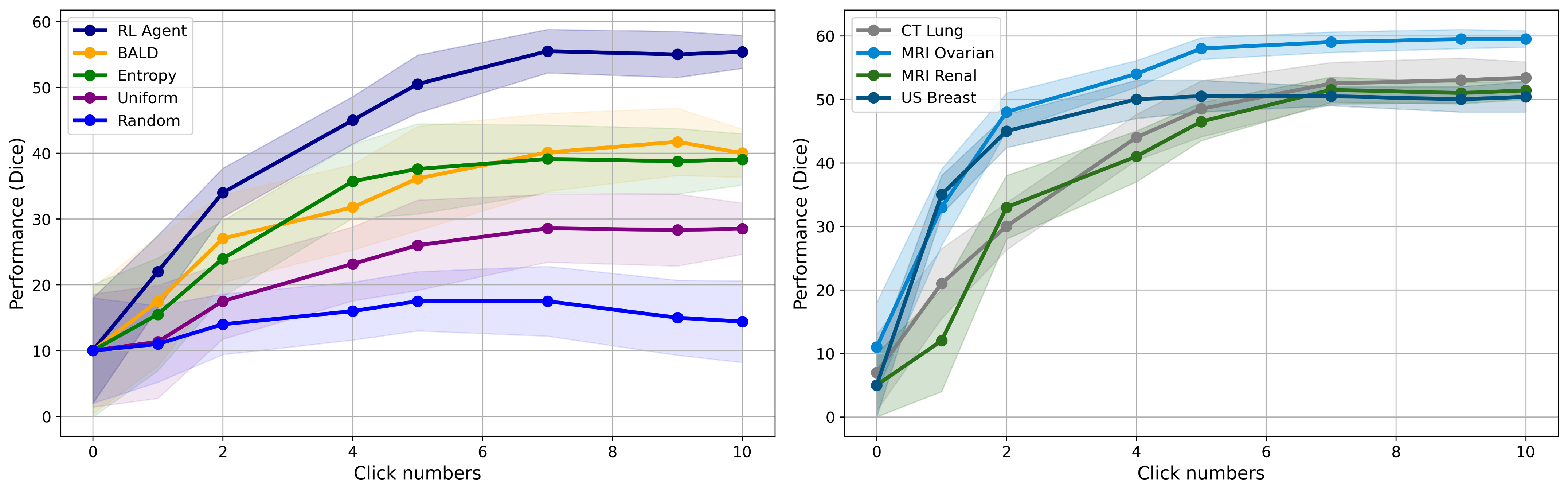}
  \caption{Reinforcement agent generated prompt masks dice performance (left) and performance for each specific dataset (right). RL Agent refers to our proposed RL framework in this paper. "BALD" refers to selected regions selected based on the maximum cumulative pixel-wise BALD from the Agent \cite{houlsby2011bayesian,houlsby2014efficient}.  "Entropy" denotes an uncertainty-based sampling agent prioritizing regions with the highest cumulative pixel-wise Shannon entropy \cite{wu2013local}. "Uniform" represents a uniform random sampling strategy \cite{kala2023autonomous}, where regions are randomly selected at each step from the available pool of regions.  Lastly, "Random" mimics real-world randomness through a random sampling process. Thirty iterations of Monte Carlo (MC) Dropout \cite{gal2016theoretically,gal2016dropout} were used to estimate uncertainty.}
  \label{fig:RL_agent_performance}
\end{figure}

\textbf{Segmentation performance:} As shown in Figure \ref{fig:RL_agent_performance}, the Dice coefficients of segmentation masks generated by the RL agent improved with an increasing number of prompt points. Our proposed RL agent significantly outperforms other RL agents and prompt selection methods, including BALD \cite{houlsby2011bayesian,houlsby2014efficient}, Entropy \cite{wu2013local}. Uniform \cite{kala2023autonomous}, and Random, as demonstrated in Figure \ref{fig:RL_agent_performance}, achieving a substantial margin of improvement. Specifically, The RL agent significantly outperformed random selection, achieving a peak Dice coefficient of 0.539 ± 0.162 compared to 0.144 ± 0.262 (P $<$ 0.05). Across the four datasets, the peak Dice coefficients were as follows: 0.595 ± 0.012 for Ovarian tumor, 0.505 ± 0.012 for Renal tumor, 0.540 ± 0.018 for Lung tumor, and 0.515 ± 0.014 for Breast tumor. These results highlight that the RL agent achieved the highest Dice coefficient for the lung tumor dataset compared to the ovarian, renal, and breast tumor datasets.
The overall performance of the RL agent was still lower than that of the radiologists, with a peak Dice coefficient of 0.539 ± 0.162 compared to 0.761 ± 0.026, as shown in Supplementary Figure \ref{fig:RL_agent_performance_supp}. 

\textbf{Segmentation time efficient:} As shown in Table \ref{tab:rl_prompt_numbers_dice}, RL agent significantly reduced segmentation time compared to radiologist manual segmentation (radiologist manual: 171.8 ± 73.9 secs, RL agent: 15.8 ± 5.3 secs, all P $<$ 0.05), saving over 156 secs per patients on average. The results highlight that the proposed RL agent consistently offers time savings over manual SAM methods across both imaging sequences in the four datasets.

\section*{Discussion}

Prompt strategies for SAM segmentation tailored to different ROI characteristics, such as size and imaging modality, remain an underexplored area. Our study investigated combinational prompt number and location strategies across imaging datasets with varying pathology, size, and modalities. In addition, we introduced a proof-of-concept RL agent that, while performing below radiologists' levels, demonstrates potential to reduce decision-making and establish prompt stop criteria for more efficient SAM-assisted segmentation workflows. 

Consistent with previous findings \cite{MAZUROWSKI2023102918,he2023adaptingsegmentmodelsam,cheng2023sammedicalimagescomprehensive, huang2024segment,yan2022sam,de2023testing}, our study showed that SAM's segmentation performance improved across all four datasets as the number of prompt points increased. However, our results also suggest that each pathology and modality has an optimal threshold for the number of prompts, beyond which additional prompts yield plateaued performance. For smaller ROIs, such as those in lung, renal, and breast tumors, optimal performance was achieved with 2-4 prompt points, as no significant improvements were observed with 5 or more points. In contrast, larger ROIs, such as ovarian tumors, continued to show performance gains with up to 5 or more points. Although previous works have noted improved SAM performance with increased ROI size \cite{he2023computer}, our results showed relatively similar performance between the datasets, especially upon reaching prompts plateau. 

Findings on the effect of point prompt location remain limited with previous work either only noting center prompt points or not specifying location at all. Trends in point location strategies also varied with ROI sizes. For smaller ROIs (renal and lung tumors), no significant differences were observed among the center, surface, and union prompt groups. However, for larger ROIs (ovarian tumors) and breast tumors, union and surface prompt groups outperformed center prompts. The inferior performance of the center prompt group may be explained by the difficulty optimizing and assessing the center of irregular lesions \cite{rafaeli2024prompt}. Differences in imaging resolution may explain variations in performance, aligning with the findings of Hao et. al. 2021\cite{sabottke2020effect} and Sabottke et. al. 2020 \cite{hao2021segmentation}, which highlight that optimal image resolution selection can significantly enhance neural network performance across various radiology-based machine learning tasks. More recently, Rafaeli et al 2024 \cite{rafaeli2024prompt}, new citation showed higher SAM segmentation performance with high resolution compared to low-resolution images. Interestingly, we did not find that our low-resolution modality dataset (ultrasound breast) had markedly worse SAM performance compared to the other high-resolution modalities (CT and MRI). However, the different patterns in prompt location strategy for breast tumor segmentation (low-resolution ultrasound) differed from renal and lung tumor segmentation (high-resolution CT and MRI), suggests the impact of resolution on prompt efficacy and warrants further investigation into tailoring prompt strategies to modality resolution and quality


Developing an effective stop criteria agent remains a critical focus for SAM-assisted segmentation. Our proposed method demonstrated significant improvements compared to other RL agents and selection methods, including BALD \cite{houlsby2011bayesian,houlsby2014efficient}, Entropy \cite{wu2013local}. Uniform \cite{kala2023autonomous}, and Random, achieving a notable performance margin. This enhancement highlights the effectiveness of our RL framework in optimizing prompts for segmentation tasks. The proof-of-concept RL agent shows a promising approach by identifying optimal points where additional prompts no longer improve performance, thereby streamlining segmentation workflows and reducing radiologists' interaction time. Future research should aim to refine RL agents further, enhancing their decision-making capabilities to closely align with human expertise and enabling dynamic prompt adjustments based on real-time feedback.

\subsection*{Conclusion}

This study highlights the importance of optimizing SAM segmentation strategies based on ROI size and imaging modality to maximize its performance. While additional prompts generally improve segmentation accuracy, our findings reveal optimal thresholds for each pathology and modality. 
Union or surface prompt locations outperformed center prompts for larger ROIs. Our proof-of-concept RL agent demonstrates potential to reduce decision time and optimize prompt strategies, paving the way toward transforming SAM from a semi-automated tool into a fully automated solution for lesion-specific segmentation in clinical workflows.

\subsection*{Limitations of the study}

A major limitation of this study lies in the challenge of surpassing human-assisted SAM performance, as such workflows already approach near-human accuracy. The critical next step is not merely to exceed human-assisted performance but to transition SAM into a fully automated tool while maintaining its high accuracy. This requires advanced strategies, such as refined RL agents, and access to larger, diverse annotated datasets to further improve decision-making capabilities and streamline segmentation workflows.

\section*{Acknowledgments}

This work was supported in part by an anonymized agency.

\newpage
\bibliography{references}

\newpage
\section*{Tables}

\renewcommand{\arraystretch}{1.2} 
\begin{table}[h!] 
\centering 
\caption{Patient characteristics across all four datasets} 
\label{tab:tumor_dataset_summary} 
\begin{tabular}{ |p{3cm}||p{2.8cm}|p{2.8cm}|p{2.8cm}|p{2.8cm}|  }
 \hline
 \multicolumn{5}{|c|}{Tumor Dataset Summary} \\
 \hline
 \textbf{Tumor Type} & \textbf{Ovarian Tumor} & \textbf{Renal Tumor} & \textbf{Lung Tumor} & \textbf{Breast Tumor} \\
 \hline
 Modality & MRI & MRI & CT & Ultrasound \\
 \hline
 No. of Patients & 15 & 15 & 20 & 40 \\
 \hline
 Years Collected & 2010--2020 & 2010--2019 & Unknown & 2019--2022 \\
 \hline
 Source & Institute A 
 & Institute B 
 & NSCLC-Radiomics \cite{aerts2014nsclc} & Breast-Lesions-USG \cite{pawlowska2024breast} \\
 \hline
 Age & 43.5 (+/- 14.5) & 61.0 (+/- 25.7) & Unknown & 18--17 \\
 \hline
 Sex (M:F) & 0:15 & 7:8 & 14:6 & 0:40 \\
 \hline
 ROI (avg, cm) & 8.9 (+/- 4.2) & 3.4 (+/- 1.6) & 3.0 (+/- 1.1) & 3.7 (+/- 1.1) \\
 \hline
\end{tabular}
\end{table}

\renewcommand{\arraystretch}{1.2} 
\begin{table}[h!]
\centering
\caption{Mean Dice Coefficients and 95th Percentile Hausdorff Distance (HD95) across all datasets for various point number strategies }
\label{tab:performance_metrics}
\begin{tabular}{|c|c|c|c|c|}
\hline
\textbf{Number of Points} & \multicolumn{4}{|c|}{\textbf{Dice Coefficient (Mean ± Standard Deviations)}} \\ 
\cline{2-5}
& \textbf{Ovarian Tumor} & \textbf{Renal Tumor} & \textbf{Lung Tumor} & \textbf{Breast Tumor} \\ 
\hline
1 & 0.272 ± 0.075 & 0.300 ± 0.057 & 0.389 ± 0.098 & 0.512 ± 0.070 \\ 
\hline
2--4 & 0.597 ± 0.129 & 0.687 ± 0.113 & 0.736 ± 0.045 & 0.780 ± 0.061 \\ 
\hline
5 or more & 0.806 ± 0.084 & 0.708 ± 0.108 & 0.761 ± 0.040 & 0.790 ± 0.062 \\ 
\hline
\multicolumn{5}{|c|}{\textbf{HD95 (mm, Mean ± Standard Deviations)}} \\ 
\hline
1 & 32.751 ± 6.179 & 23.616 ± 5.850 & 14.596 ± 3.004 & 26.992 ± 5.157 \\ 
\hline
2--4 & 17.412 ± 6.327 & 4.703 ± 2.415 & 6.930 ± 2.262 & 12.662 ± 4.460 \\ 
\hline
5 or more & 8.450 ± 4.622 & 4.740 ± 2.630 & 5.815 ± 1.823 & 10.859 ± 3.862 \\ 
\hline
\end{tabular}
\end{table}

\newpage
\renewcommand{\arraystretch}{1.2} 
\begin{table}[h!]
\centering
\caption{Dice Coefficients and the 95th Percentile Hausdorff Distance (HD95) across all datasets for various point location trategies.}
\label{tab:point_location_metrics}
\begin{tabular}{|c|c|c|c|c|}
\hline
\textbf{Point Location} & \multicolumn{4}{|c|}{\textbf{Dice Coefficient (Mean ± Standard Deviations)}} \\ 
\cline{2-5}
& \textbf{Ovarian Tumor} & \textbf{Renal Tumor} & \textbf{Lung Tumor} & \textbf{Breast Tumor} \\ 
\hline
Center & 0.491 ± 0.246 & 0.563 ± 0.249 & 0.628 ± 0.156 & 0.650 ± 0.176 \\ 
\hline
Surface & 0.580 ± 0.272 & 0.548 ± 0.222 & 0.625 ± 0.228 & 0.708 ± 0.185 \\ 
\hline
Union & 0.604 ± 0.284 & 0.583 ± 0.245 & 0.633 ± 0.245 & 0.724 ± 0.189 \\ 
\hline
\multicolumn{5}{|c|}{\textbf{HD95 (mm, Mean ± Standard Deviations)}} \\ 
\hline
Center & 22.559 ± 12.613 & 11.470 ± 11.107 & 9.266 ± 5.019 & 19.359 ± 9.860 \\ 
\hline
Surface & 18.654 ± 12.713 & 11.038 ± 10.986 & 8.363 ± 4.545 & 15.714 ± 10.016 \\ 
\hline
Union & 17.402 ± 12.400 & 10.552 ± 11.007 & 9.713 ± 5.640 & 15.439 ± 9.658 \\ 
\hline
\end{tabular}
\end{table}

\renewcommand{\arraystretch}{1.2} 
\begin{table}[h!]
\centering
\caption{Mean annotation time across all datasets with different prompt numbers.}
\label{tab:rl_prompt_numbers_dice}
\begin{tabular}{|c|c|c|c|c|}
\hline
\textbf{Time (secs)} & \textbf{MRI Ovarian} & \textbf{MRI Renal} & \textbf{CT Lung} & \textbf{US Breast} \\ 
\hline
\textbf{Radiologist} & 237.0 ± 72.0 & 150.6 ± 68.4 & 286.8 ± 109.2 & 12.6 ± 5.4 \\
\hline
\textbf{RL Agent} & 21.8 ± 1.8 & 10.8 ± 9.2 & 29.0 ± 4.8 & 1.4 ± 1.2 \\ 
\hline
\end{tabular}
\end{table}

\newpage
\section*{Figures}

\begin{figure}[H]
  \centering
  \includegraphics[width=0.7\textwidth]{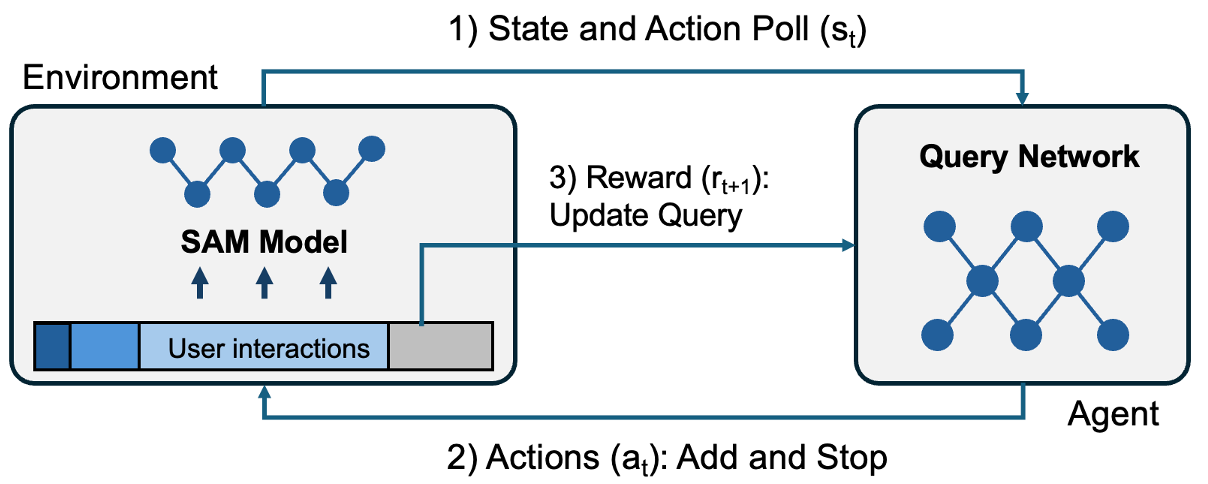}
  \caption{Reinforcement agent architecture and workflow. The agent perceives the environment, selects an action, and updates its policy iteratively to maximize cumulative rewards.}
  \label{fig:rl_workflow}
\end{figure}

\begin{figure}[H]
  \centering
  \includegraphics[width=0.85\textwidth]{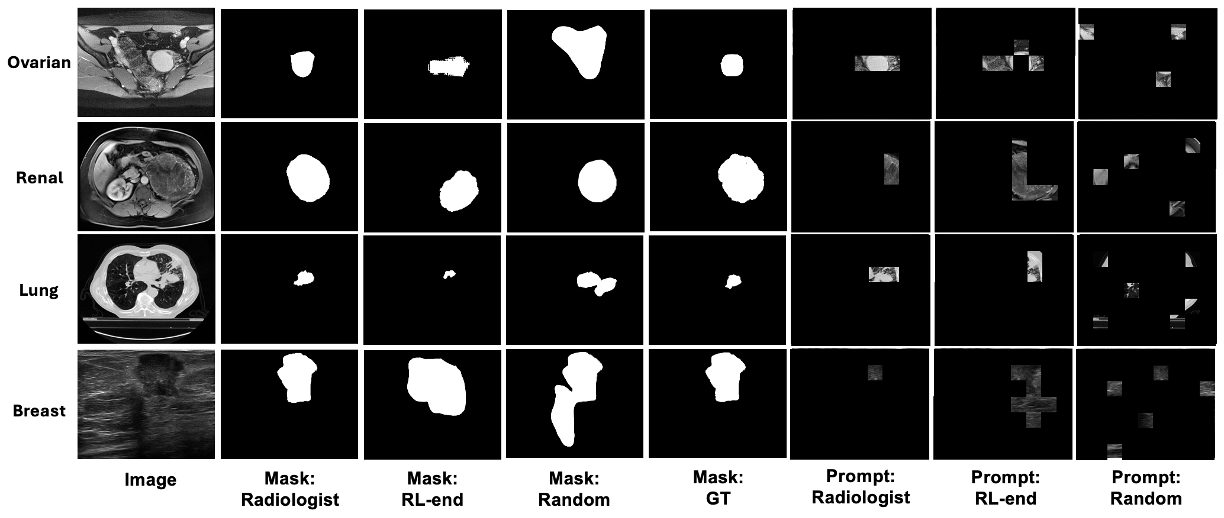}
  \caption{Reinforcement agent generated prompt masks (left) and prompts (right).}
  \label{fig:RL_mask_performance}
\end{figure}

\newpage
\section*{Supplemental information}

\subsection*{Data and code availability}
Our source code, developed based on RALIS\cite{casanova2020reinforced}. We redefined the state representation, action representation, and DQN \cite{mnih2013playing} to adapt them to our SAM project.

\subsection*{Technical details of RL agent}
We propose to model the SAM segmentation task as a sequential decision-making problem, illustrated in Figure \ref{fig:rl_workflow}. An agent is built to interact with the visual features or kinematic prompt sequence, i.e., the environment. At each time step, the agent observes the state of the environment and selects an action (i.e., a next prompts location) accordingly, which in turn affects the environment. The agent earns a numerical reward for each action and updates its policy to maximize future rewards. Following the standard paradigm of reinforcement learning, we formalize the task as a Markov Decision Process, $\mathcal{M} := (\mathcal{S}, \mathcal{A}, \mathcal{P}, \mathcal{R}, \gamma)$, 
where $\mathcal{S}$ is a finite set of states, $\mathcal{A}$ is a finite set of actions, $\mathcal{P} : \mathcal{S} \times \mathcal{A} \times \mathcal{S} \to [0, 1]$ denotes state transition probabilities, 
$\mathcal{R} : \mathcal{S} \times \mathcal{A} \to \mathcal{R}$ denotes a reward function for each action performed in a certain state, and $\gamma \in [0, 1]$ is the discount factor balancing between immediate and long-term reward. The agent learns from experience to optimize its policy $\pi : \mathcal{S} \times \mathcal{A} \to [0, 1]$, which is stochastic in general. 
The goal is to maximize the discounted future reward accumulated from the current time step to the end of the learning episode.

\subsubsection*{State representation} We would like to use the state of the SAM segmentation output as the MDP state. The state representation is built by concatenating three key features that capture different aspects of the segmentation task: entropy between the background and foreground, the gradient vector field along the segmentation mask boundary, and the class distribution, which is formalized below:
\begin{equation}
    s_t(X_n) := \left(s^t_{\text{H}}, s^{t}_{\text{argmax}}, s^{t}_{\text{gradient}}\right)
\end{equation}
where $s^t_{\text{H}}$, the entropy feature measures uncertainty in distinguishing between background and foreground, highlighting ambiguous areas at the current frame. $s^{t}_{\text{argmax}}$, The gradient vector field captures edge and contour information, helping refine boundaries, while the class distribution shows the model’s confidence in pixel classifications. $s^{t}_{\text{gradient}}$ is the distribution feature. By combining these features across all regions, the state representation provides a comprehensive view of segmentation quality, guiding further refinement or labeling.

\begin{figure}[H]
  \centering
  \includegraphics[width=0.7\textwidth]{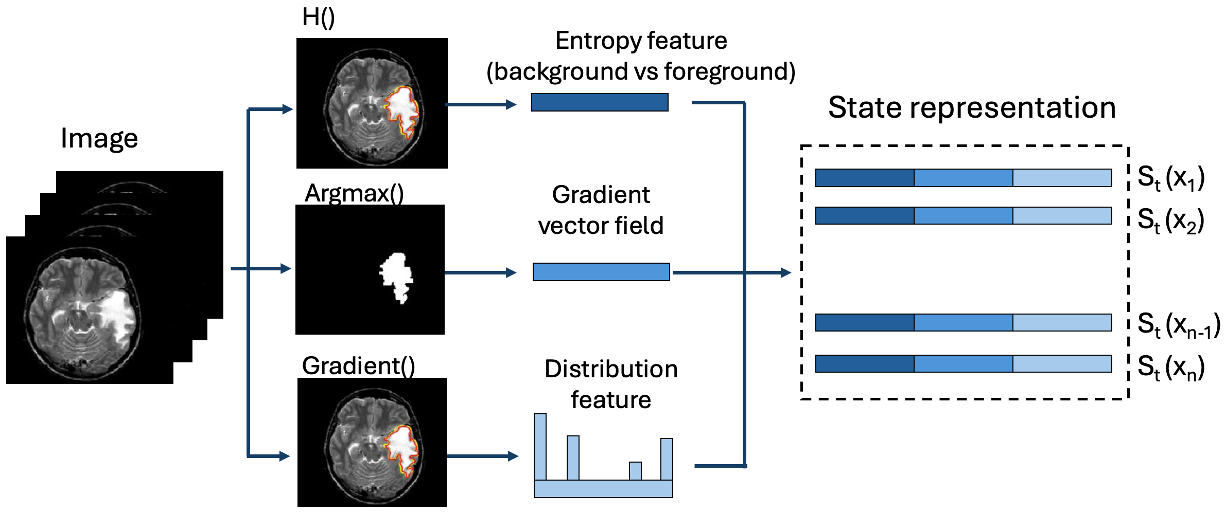}
  \caption{Explanation of the state representation, which is constructed as a concatenation of three features: one based on the entropy between the background and foreground, another based on the gradient vector field along the segmentation mask boundary, and the third on the class distribution. The final state is the concatenation of these features for all regions.}
  \label{fig:sm_state_repre}
\end{figure}

\subsubsection*{Action representation} The action representation defines an action pool from which new prompts can be selected. It is composed of a concatenation of five features: entropy-based features, the gradient vector field, a distribution feature, and two KL divergence features that compare each region with both the labeled and unlabeled sets. Specifically, the KL divergence is calculated between the class distributions of the prediction map (estimated as normalized counts of predicted pixels per category) and the class distributions of both labeled and unlabeled regions (using ground truth annotations and network predictions, respectively). For the labeled set, a KL divergence score is computed between each labeled region's class distribution and that of region. These KL divergence values can be summarized by either taking the maximum value or summing them together.

\begin{figure}[H]
  \centering
  \includegraphics[width=0.7\textwidth]{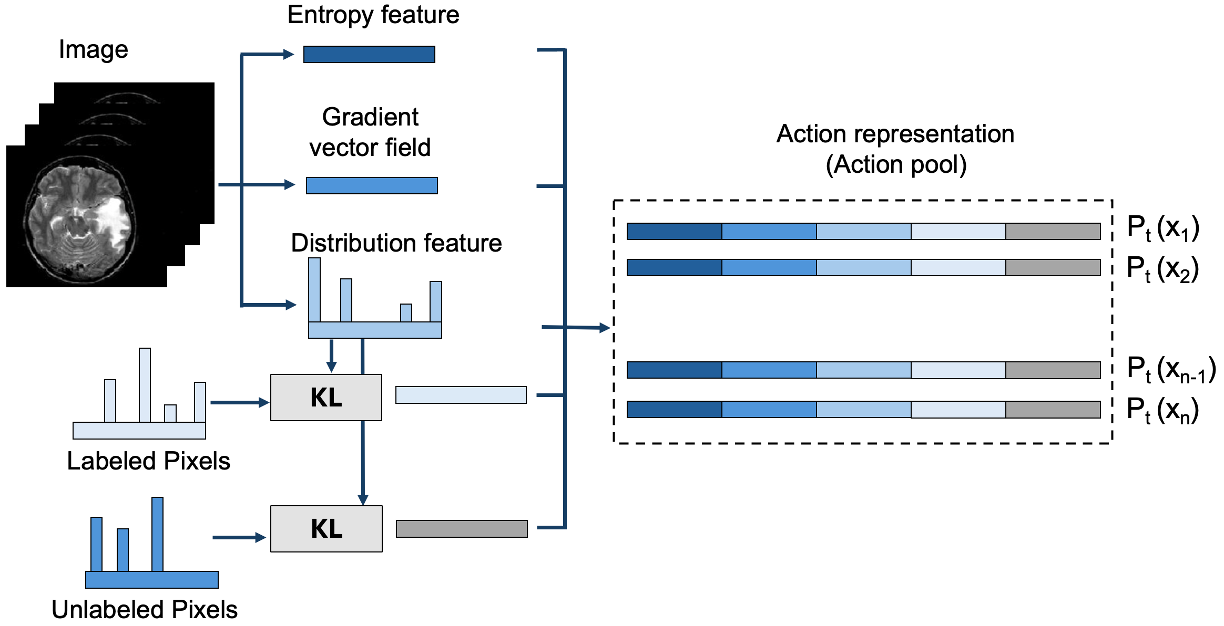}
  \caption{Explanation of the Action representation, which is represented as a concatenation of five features: entropy-based features, gradient vector field, distribution feature, and two KL divergence distributions, comparing each region with the labeled and unlabeled set.}
  \label{fig:sm_action_repre}
\end{figure}

\subsubsection*{Experimental setup and training details}
The query network, shown in Supplementary Figure \ref{fig:dqn}, consists of two branches: one for computing state features and another for computing action features, which are fused at the end. Each layer includes Batch Normalization, ReLU activation, and a fully connected layer. The state and action branches contain 4 and 3 layers, respectively, with a final layer that combines them to obtain the global features. These features are gated using a sigmoid function controlled by the KL divergence in the action representation. The weights are updated at each step of the prompt interactive loop. We train both networks using stochastic gradient descent with momentum. The learning rates for both the segmentation and query networks are set to 10$^{-4}$ for all datasets. Weight decay is set to 10$^{-4}$ for the segmentation network and 10$^{-3}$ for the query network. 

\begin{figure}[H]
  \centering
  \includegraphics[width=0.4\textwidth]{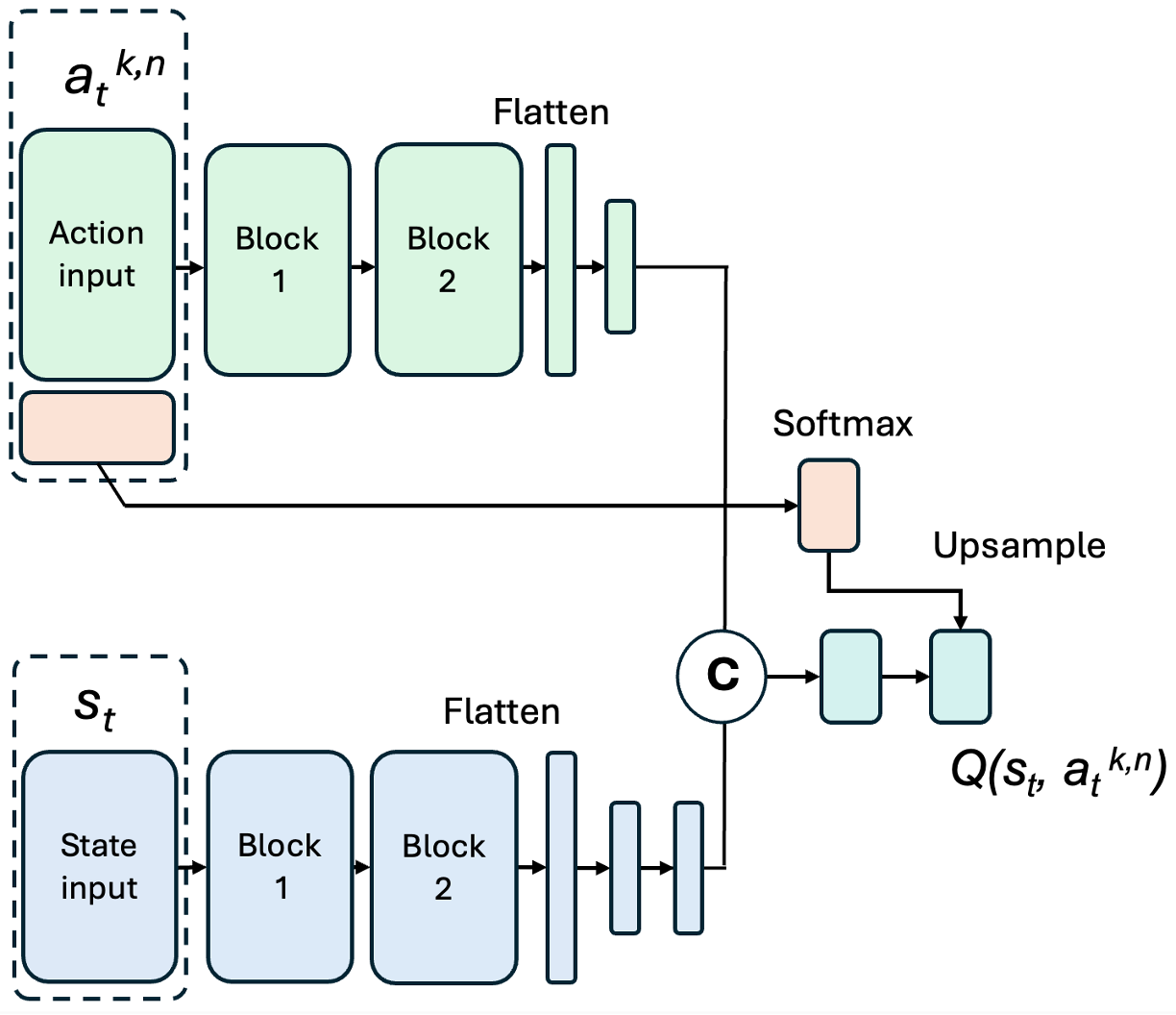}
  \caption{The DQN processes a state representation and an action representation for a potential action. The blue block represents the number of state and action features (entropy-based features, gradient vector field, distribution feature), and the organ block represents the features for the KL divergence distributions. Both representations are computed separately using layers with Batch Normalization, ReLU activation, and fully connected layers. The feature vectors are then flattened, concatenated, and passed through a final linear layer to produce a scalar score. The Q-values are determined as a gated score, where the gate is modulated by the KL distance features from the action representation.}
  \label{fig:dqn}
\end{figure}

\renewcommand{\arraystretch}{1.2} 
\begin{table}[h!]
\centering
\caption{Quantitative Dice Summary: Reinforcement Agent Performance for Each Specific Dataset.}
\label{tab:reinforcement_agent_performance}
\begin{tabular}{|c|c|c|c|c|}
\hline
\textbf{Prompt Number} & \multicolumn{4}{|c|}{\textbf{Dice Coefficient (Mean ± Standard Deviations)}} \\ 
\cline{2-5}
& \textbf{CT Lung} & \textbf{MRI Ovarian} & \textbf{MRI Renal} & \textbf{US Breast} \\ 
\hline
0  & 0.0791 ± 0.028 & 0.119 ± 0.032 & 0.056 ± 0.027 & 0.055 ± 0.032 \\ 
\hline
1  & 0.218 ± 0.024 & 0.339 ± 0.025 & 0.351 ± 0.017 & 0.123 ± 0.028 \\ 
\hline
2  & 0.310 ± 0.019 & 0.481 ± 0.017 & 0.458 ± 0.016 & 0.334 ± 0.022 \\ 
\hline
4  & 0.449 ± 0.019 & 0.549 ± 0.015 & 0.502 ± 0.017 & 0.411 ± 0.021 \\ 
\hline
5  & 0.485 ± 0.021 & 0.588 ± 0.013 & 0.505 ± 0.016 & 0.465 ± 0.017 \\ 
\hline
7  & 0.525 ± 0.018 & 0.594 ± 0.013 & 0.505 ± 0.012 & 0.515 ± 0.014 \\ 
\hline
9  & 0.534 ± 0.019 & 0.595 ± 0.012 & 0.502 ± 0.014 & 0.513 ± 0.013 \\ 
\hline
10 & 0.534 ± 0.016 & 0.595 ± 0.011 & 0.504 ± 0.016 & 0.514 ± 0.012 \\ 
\hline
\end{tabular}
\end{table}



\begin{figure}[H]
  \centering
  \includegraphics[width=0.85\textwidth]{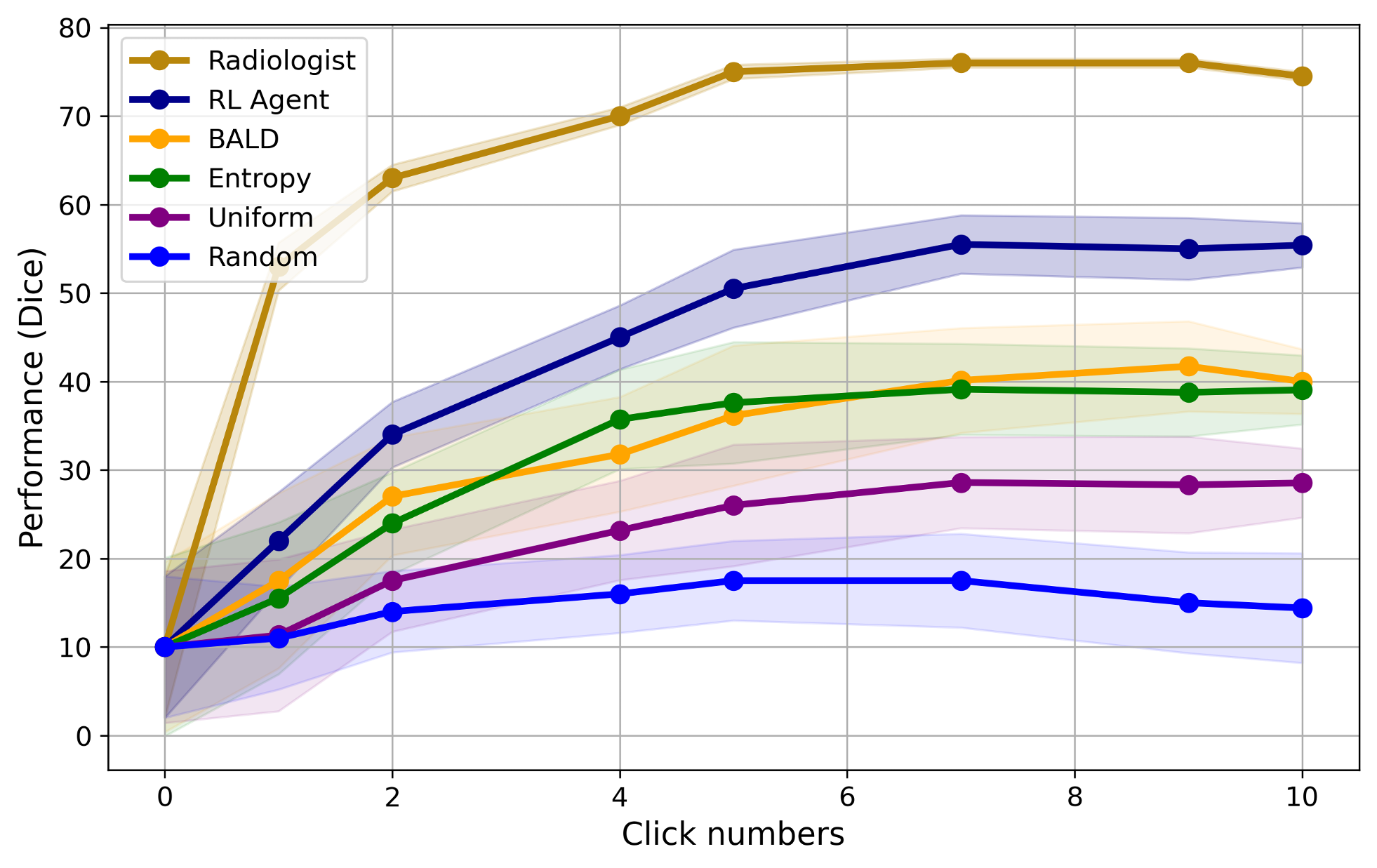}
  \caption{Reinforcement agent generated prompt masks dice performance. Radiologist refers to the manual segmentation performance achieved using SAM by two radiologists. RL Agent refers to our proposed RL framework in this paper. BALD refers to selected regions based on the maximum cumulative pixel-wise BALD metric \cite{houlsby2011bayesian,houlsby2014efficient}. Entropy is an uncertainty-based sampling method that prioritizes regions with the highest cumulative pixel-wise Shannon entropy \cite{wu2013local}. Uniform refers to a uniform random sampling strategy \cite{kala2023autonomous}, where regions to label are selected at each step from all available regions in the pool.  Lastly, Random refers to a random sampling process designed to mimic real-world randomness.}
  \label{fig:RL_agent_performance_supp}
\end{figure}

\section*{Declaration of interests}

The authors declare no competing interests.

\end{document}